\documentclass[twocolumn,aps,showpacs,preprintnumbers,amsmath,amssymb]{revtex4}
\usepackage{graphicx}

\begin{document}

\title{Quantum interferences from cross talk in $J=1/2\leftrightarrow J=1/2$ transitions}

\author{Shubhrangshu Dasgupta}
\affiliation{Physical Research Laboratory, Navrangpura, Ahmedabad
- 380 009, India}
\date{\today}

\begin{abstract}
We consider the possibility of a control field opening up multiple
pathways and thereby leading to new interference and coherence
effects. We illustrate the idea by considering the
$J=1/2\leftrightarrow J=1/2$ transition. As a result of the additional
pathways, we show the possibilities of nonzero refractive index
without absorption and gain without inversion. We explain these
results in terms of the coherence produced by the opening of
an extra pathway.
\end{abstract}


\maketitle

\section{\label{sec:intro}Introduction}

The usage of a coherent field to control the optical properties of
a medium has led to many remarkable results such as enhanced
nonlinear optical effects \cite{spt,harris:eit},
electromagnetically induced transparency (EIT) \cite{harris:pt},
lasing without inversion \cite{lwi,lwi:rev,gsa1:lwi}, ultraslow
light \cite{hau,scully,budker}, storage and revival of optical
pulses \cite{lukin} and many others
\cite{index1,index:expt,Zhu0304,DengPRL}. Most of these effects
rely on quantum interferences which are created by the application
of a coherent field. The coherent field opens up a new channel for
the process under consideration. The application of the coherent
field gives us considerable flexibility as by changing the
strength and the frequency of the field we can obtain a variety of
control on the optical properties of the medium. The Zeeman
degeneracy of the atomic levels adds to this flexibility as the
coherent field may open up more than one pathway thereby leading
to new interferences.

In this paper, we consider  a specific atomic energy level scheme
consisting of $J=1/2\leftrightarrow J=1/2$ transitions. We show
how a control field opens up more than one pathway and how this
leads to new coherent effects. We report interesting results on
gain without inversion and on the production of the refractive
index without absorption \cite{footnote1}.

The organization of the paper is as follows. In Sec. II, we
describe the atomic configuration in details. We put forward all
the relevant equations and the steady state solutions. In Sec.
III, we discuss the absorption and dispersion profiles of the
probe field. In Sec. IV, we show how new features arise in these
profiles as effects of the new coherence.

\begin{figure}
\scalebox{0.5}{\includegraphics{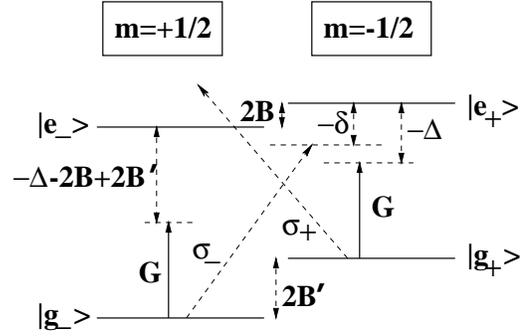}}
\caption{\label{chap5fig1}Relevant configuration of a four-level
atom. A dc magnetic field removes the degeneracy of the excited
sublevels $|e_\pm\rangle$  and the ground sublevels
$|g_\pm\rangle$. The magnetic field makes the system anisotropic.
The corresponding Zeeman separations are $2B$ and $2B'$,
respectively. The $\sigma_\pm$ components (with Rabi frequencies
$2g_\pm$) of the $\hat{x}$-polarized probe field interact with the
$|e_\mp\rangle\leftrightarrow |g_\pm\rangle$ transitions. Here
$2G$ is the Rabi frequency of the control field, interacting with
the $|e_\pm\rangle\leftrightarrow |g_\pm\rangle$ transitions. Here
$\delta$ and $\Delta$ are the detunings of the probe and control
fields.}
\end{figure}

\begin{figure}
\scalebox{0.4}{\includegraphics{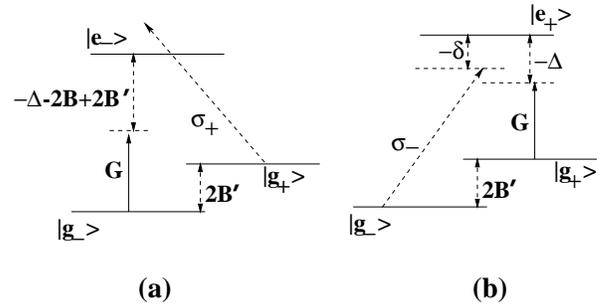}}
\caption{\label{fignew}Two $\Lambda$ systems, equivalent to the
configuration shown in Fig.~\ref{chap5fig1}. These two systems
talk to each other as both the systems are driven by the same
control field $G$.}
\end{figure}

\section{\label{sec:model}Model configuration}
We consider the $J=1/2\leftrightarrow J=1/2$ transition in alkali
atoms, as shown in Fig.~\ref{chap5fig1}. This kind of
configuration is important for studying the effects of cross talk
among different transitions \cite{brown}. Note that cross talk in
a $\Lambda$-system can lead to gain in the transmission of the
probe field \cite{menon}. Similar configuration with degenerate
sublevels also leads to electromagnetically induced absorption
while interacting with control and probe fields with the same
polarization \cite{goren}. We apply a dc magnetic field to remove
the degeneracy of the excited and the ground states. In general,
the Zeeman separation $2B$ of the excited magnetic sublevels
$m_e=\pm 1/2 (\equiv |e_\mp\rangle)$ is not the same as the Zeeman
separation $2B'$ of the ground levels $(\equiv |g_\mp\rangle)$,
due to difference in the Land\'e $g$-factors in these manifolds.
For example, in $^{39}$K atom, $B'=3B$, where $B=\mu_Bg_eM/\hbar$
($\mu_B$ is the Bohr magneton) and $g_e=2/3$ and $g_g=2$ are the
Land\'e $g$-factors of the excited and the ground sublevels.

We apply a weak $\hat{x}$ polarized field $\vec{E}_p=\hat{x}{\cal
E}_pe^{-i\omega_1 t}+\textrm{c.c.}$ to probe the properties of the
atom, where $\omega_1$ is the angular frequency of the field. The
$\sigma_\pm$ components of this probe field interact with the
$|e_\mp\rangle\leftrightarrow |g_\pm\rangle$ transitions. The Rabi
frequencies are given by $2g_\pm=2(\vec{d}_{e_\mp
g_\pm}.\hat{x}{\cal E}_p)/\hbar$, where $\vec{d}_{ij}$ is the
electric dipole moment between the levels $|i\rangle$ and
$|j\rangle$. We also apply a strong $\pi$-polarized control field
\begin{equation}
\vec{E}_c=\vec{\cal E}_ce^{-i\omega_2 t}+\textrm{c.c.}\;,
\end{equation}
which interacts with the $|e_\pm\rangle\leftrightarrow
|g_\pm\rangle$ transitions. We assume that the corresponding Rabi
frequencies $2\vec{d}_{e_\pm g_\pm}.\vec{\cal E}_c/\hbar$ are
equal to $2G$. We emphasize that the system of
Fig.~\ref{chap5fig1} can be visualized as two $\Lambda$ systems
(as shown in Fig.~\ref{fignew}), which talk to each other, as the
same control field $G$ drives both the transitions
$|e_\pm\rangle\leftrightarrow |g_\pm\rangle$.

The interaction Hamiltonian of this system in dipole approximation
is
\begin{eqnarray}
\frac{H}{\hbar}&=&\left[(\omega_{e_-g_-}|e_-\rangle\langle e_-|+\omega_{e_+g_-}|e_+\rangle\langle e_+|+\omega_{g_+g_-}|g_+\rangle\langle g_+|)\right.\nonumber\\
&-&(\vec{d}_{e_-g_+}|e_-\rangle\langle g_+|+\vec{d}_{e_+g_-}|e_+\rangle\langle g_-|+\textrm{h.c.}).\vec{E}_p\nonumber\\
&-&\left.(\vec{d}_{e_+g_+}|e_+\rangle\langle g_+|+\vec{d}_{e_-g_-}|e_-\rangle\langle g_-|+\textrm{h.c.}).\vec{E}_c\right]\;.
\label{hamil}
\end{eqnarray}
Here the zero of energy is defined at the level $|g_-\rangle$ and $\hbar\omega_{\alpha\beta}$
is the energy difference between the levels $|\alpha\rangle$ and $|\beta\rangle$.

We consider the natural decay terms in our analysis and hence invoke the
density matrix formalism to find the following equations for different density
matrix elements:
\begin{widetext}
\begin{eqnarray}
\dot{\tilde{\rho}}_{e_+g_-}&=&-\left[i(-\Delta +2B')+\Gamma_{e_+g_-}\right]\tilde{\rho}_{e_+g_-}+i\left[g_-e^{-i\omega_{12} t}(\tilde{\rho}_{g_-g_-}-\tilde{\rho}_{e_+e_+})+G\tilde{\rho}_{g_+g_-}-G\tilde{\rho}_{e_+e_-}\right]\;,\nonumber\\
\dot{\tilde{\rho}}_{e_-g_+}&=&-\left[i(-\Delta -2B)+\Gamma_{e_-g_+}\right]\tilde{\rho}_{e_-g_+}+i\left[g_+e^{-i\omega_{12} t}(\tilde{\rho}_{g_+g_+}-\tilde{\rho}_{e_-e_-})+G\tilde{\rho}_{g_-g_+}-G\tilde{\rho}_{e_-e_+}\right]\;,\nonumber\\
\dot{\tilde{\rho}}_{e_+g_+}&=&-\left[-i\Delta +\Gamma_{e_+g_+}\right]\tilde{\rho}_{e_+g_+}+i\left[G(\tilde{\rho}_{g_+g_+}-\tilde{\rho}_{e_+e_+})+g_-e^{-i\omega_{12}t}\tilde{\rho}_{g_-g_+}-g_+e^{-i\omega_{12} t}\tilde{\rho}_{e_+e_-}\right]\;,\nonumber\\
\dot{\tilde{\rho}}_{e_-g_-}&=&-\left[i(-\Delta -2B+2B')+\Gamma_{e_-g_-}\right]\tilde{\rho}_{e_-g_-}+i\left[g_+e^{-i\omega_{12} t}\tilde{\rho}_{g_+g_-}-g_-e^{-i\omega_{12} t}\tilde{\rho}_{e_-e_+}+G(\tilde{\rho}_{g_-g_-}-\tilde{\rho}_{e_-e_-})\right]\;,\nonumber\\
\label{doteq}\dot{\tilde{\rho}}_{g_+g_-}&=&-(2iB'+\Gamma_{g_+g_-})\tilde{\rho}_{g_+g_-}+i\left[G^*\tilde{\rho}_{e_+g_-}-G\tilde{\rho}_{g_+e_-}+g_+^*e^{i\omega_{12} t}\tilde{\rho}_{e_-g_-}-g_-e^{-i\omega_{12} t}\tilde{\rho}_{g_+e_+}\right]\;,\\
\dot{\tilde{\rho}}_{e_+e_-}&=&-(2iB+\Gamma_{e_+e_-})\tilde{\rho}_{e_+e_-}-i\left[G^*\tilde{\rho}_{e_+g_-}-G\tilde{\rho}_{g_+e_-}+g_+^*e^{i\omega_{12} t}\tilde{\rho}_{e_+g_+}-g_-e^{-i\omega_{12} t}\tilde{\rho}_{g_-e_-}\right]\;,\nonumber\\
\dot{\tilde{\rho}}_{g_-g_-}&=&\gamma_{g_-e_-}\tilde{\rho}_{e_-e_-}+\gamma_{g_-e_+}\tilde{\rho}_{e_+e_+}+i\left[g_-^*e^{i\omega_{12} t}\tilde{\rho}_{e_+g_-}+G^*\tilde{\rho}_{e_-g_-}-\textrm{h.c.}\right]\;,\nonumber\\
\dot{\tilde{\rho}}_{e_-e_-}&=&-(\gamma_{g_-e_-}+\gamma_{g_+e_-})\tilde{\rho}_{e_-e_-}+i\left[g_+e^{-i\omega_{12} t}\tilde{\rho}_{g_+e_-}+G\tilde{\rho}_{g_-e_-}-\textrm{h.c.}\right]\;,\nonumber\\
\dot{\tilde{\rho}}_{e_+e_+}&=&-(\gamma_{g_-e_+}+\gamma_{g_+e_+})\tilde{\rho}_{e_+e_+}+i\left[g_-e^{-i\omega_{12}
t}\tilde{\rho}_{g_-e_+}+G\tilde{\rho}_{g_+e_+}-\textrm{h.c.}\right]\;,\nonumber
\end{eqnarray}
\end{widetext}
where $\Delta =\omega_2-\omega_{e_+g_+}$ is the detuning of the
control field from the transition $|e_+\rangle\leftrightarrow
|g_+\rangle$, $\delta=\omega_1-\omega_{e_+g_-}$ is the detuning of
the $\sigma_-$ component of the probe field from the transition
$|e_+\rangle\leftrightarrow |g_-\rangle$,
$\omega_{12}=\omega_1-\omega_2=\delta-\Delta+2B'$ is the
difference between frequencies of the probe and control fields,
$\gamma_{\alpha\beta}$ is the spontaneous emission rate from the
level $|\beta\rangle$ to $|\alpha\rangle$,
$\Gamma_{\alpha\beta}=\frac{1}{2}\sum_k(\gamma_{k\alpha}+\gamma_{k\beta})$
is the dephasing rate of the coherence between the levels
$|\alpha\rangle$ and $|\beta\rangle$. Here onwards, we assume that
$\gamma_{g_+e_-}=\gamma_{g_-e_+}=\gamma_1$ and
$\gamma_{g_+e_+}=\gamma_{g_-e_-}=\gamma_2$ without loss of
generality, so that
$\Gamma_{e_+g_\pm}=\Gamma_{e_-g_\pm}=\Gamma=(\gamma_1+\gamma_2)/2$,
$\Gamma_{e_+e_-}=\gamma_1+\gamma_2$, and $\Gamma_{g_+g_-}=0$
\cite{note}.
\begin{figure*}
\begin{tabular}{cc}
\scalebox{0.37}{\includegraphics{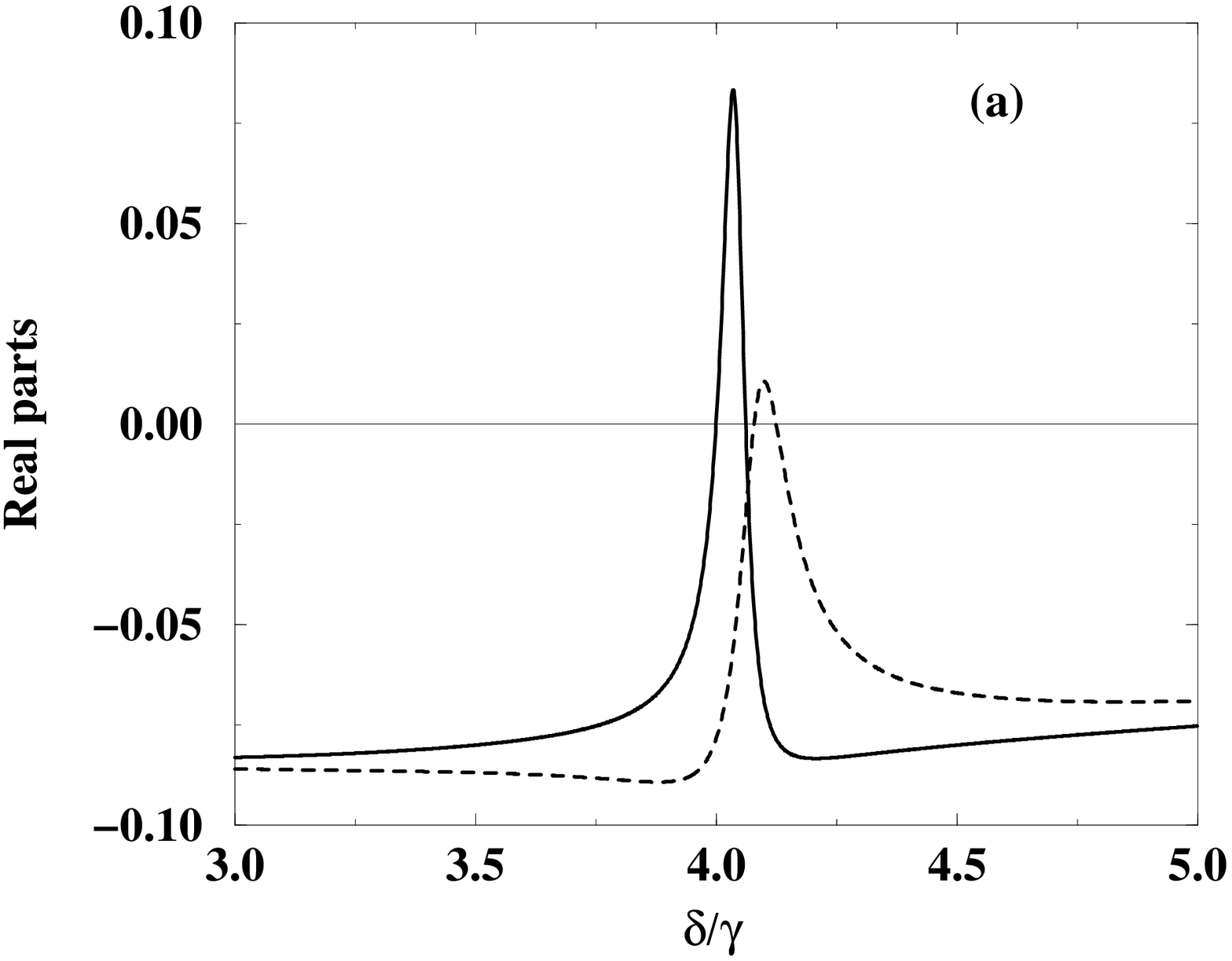}}&
\scalebox{0.37}{\includegraphics{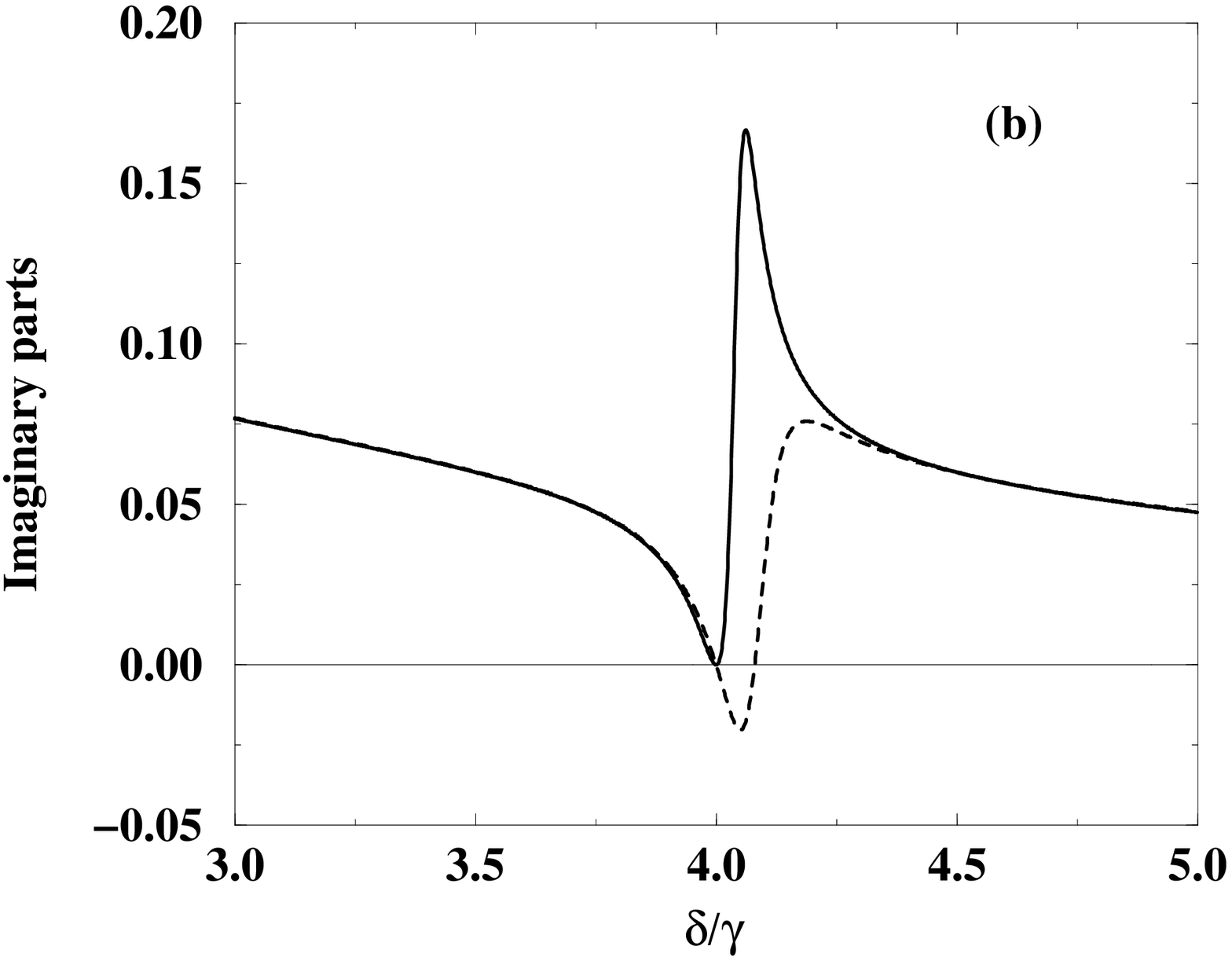}}
\end{tabular}
\caption{\label{chap5fig2}Variation of (a) real and (b) imaginary
parts of $\chi_-$ in units of $N|\vec{d}_{e_+g_-}|^2/\hbar\gamma$
with the detuning $\delta$ for the $\Lambda$ system as shown in
Fig.~\ref{fignew}(b) (solid line) and for the system shown in
Fig.~\ref{chap5fig1} (dashed line). The parameters chosen here are
$G=0.5\gamma$, $\Delta=B'-B$, $B=2\gamma$, $B'=3B$,
$\gamma_1=4\gamma$, $\gamma_2=2\gamma$. The parameter $\gamma$ is
defined as $\gamma=A/12$, taking all the decays from
$|e_\pm\rangle$ to $|g_\pm\rangle$ levels into account,
$A=2\pi\times 6.079$ MHz being the Einstein's A-coefficient in
$^{39}$K.}
\end{figure*}
To obtain the above equations, we have applied the rotating wave
approximation to neglect the highly oscillating terms. The
transformed matrix elements are given by $\tilde{\rho}_{e_\pm
g_\mp}=\rho_{e_\pm g_\mp}e^{i\omega_2 t}$ and $\tilde{\rho}_{e_\pm
g_\pm}=\rho_{e_\pm g_\pm}e^{i\omega_2 t}$, whereas the other
elements remain unchanged. Before solving Eqs. (\ref{doteq}), let
us first analyze two different cases.

Case I: We consider Fig.~\ref{fignew}(b). If the probe field
$\sigma_-$ is absent, all the populations from the level
$|g_+\rangle$ would be optically pumped to the state $|g_-\rangle$
by the action of the control field $G$. Thus, the level
$|g_-\rangle$ would be the steady state. When the probe field is
on, population transfer from the level $|g_-\rangle$ to
$|g_+\rangle$ would occur via the following pathway: absorption
from the $\sigma_-$ component followed by the emission in the
control field in the $|e_+\rangle\rightarrow |g_+\rangle$
transition. The relevant susceptibility of the probe field would
be the same as that in case of a $\Lambda$ system. We provide the
corresponding expression at the end of this section.

Case II: We consider Fig.~\ref{chap5fig1}. If both the
$\sigma_\pm$ components are absent, then in the steady state, the
population gets distributed in four levels depending upon the
detunings and the Rabi frequencies of the control fields. In this
case, when both the $\sigma_\pm$ components are switched on, an
{\it additional\/} pathway for population transfer from the level
$|g_-\rangle$ to $|g_+\rangle$ would arise, in addition to the one
described in the Case I. This pathway can be described as follows:
absorption from the control field in $|g_-\rangle\rightarrow
|e_-\rangle$ transition followed by the emission in the $\sigma_+$
component. Interference of these two pathways (i.e., cross talk
between two $\Lambda$ systems in Fig.~\ref{fignew}) leads to new
coherence in the four-level system of Fig.~\ref{chap5fig1}, which
would not arise in a $\Lambda$-system (Case I). Later we show that
all the new features described in this paper can be attributed to
this coherence. Note that coherences have been recognized as the
major source of newer effects in multilevel systems
\cite{harris:pt,lwi:rev,gsa1:lwi}.

For the Case II, the steady state solutions of the equations
(\ref{doteq}) can be found by expanding the density matrix
elements in terms of the harmonics of $\omega_{12}$ as
\begin{eqnarray}
\tilde{\rho}_{\alpha\beta}&=&\tilde{\rho}_{\alpha\beta}^{(0)}+g_-e^{-i\omega_{12} t}\tilde{\rho}_{\alpha\beta}^{'(-1)}+g_-^*e^{i\omega_{12} t}\tilde{\rho}_{\alpha\beta}^{''(-1)}\nonumber\\
&&+g_+e^{-i\omega_{12} t}\tilde{\rho}_{\alpha\beta}^{'(+1)}+g_+^*e^{i\omega_{12} t}\tilde{\rho}_{\alpha\beta}^{''(+1)}\;.
\end{eqnarray}
Thus, we obtain a set of algebraic equations for
$\tilde{\rho}_{\alpha\beta}^{(r)}$'s. Solving them, we find the
following zeroth order population terms (i.e., when both the
$\sigma_\pm$ components are absent):
\begin{subequations}
\label{popul}
\begin{eqnarray}
\tilde{\rho}_{e_\pm e_\pm}^{(0)}&=&\frac{xy}{Q}\;,\\
\tilde{\rho}_{g_-g_-}^{(0)}&=&\frac{y}{Q}(\gamma_1+\gamma_2+x)\;,\\
\tilde{\rho}_{g_+g_+}^{(0)}&=&\frac{x}{Q}(\gamma_1+\gamma_2+y)\;,
\end{eqnarray}
\end{subequations}
where
\begin{eqnarray}
&&x=\frac{2|G|^2\Gamma}{|d|^2}\;;~y=\frac{2|G|^2\Gamma}{|c|^2}\;,\nonumber\\
&&Q=(x+y)(\gamma_1+\gamma_2)+4xy\;,\nonumber\\
&&c=i\Delta+\Gamma\;,\\
&&d=-i(-\Delta-2B+2B')+\Gamma\;.\nonumber
\end{eqnarray}
Clearly, the population is distributed among the levels
$|e_\pm\rangle$ and $|g_\pm\rangle$. This is due to optical
pumping as we have discussed earlier.
\begin{figure*}
\begin{tabular}{cc}
\scalebox{0.37}{\includegraphics{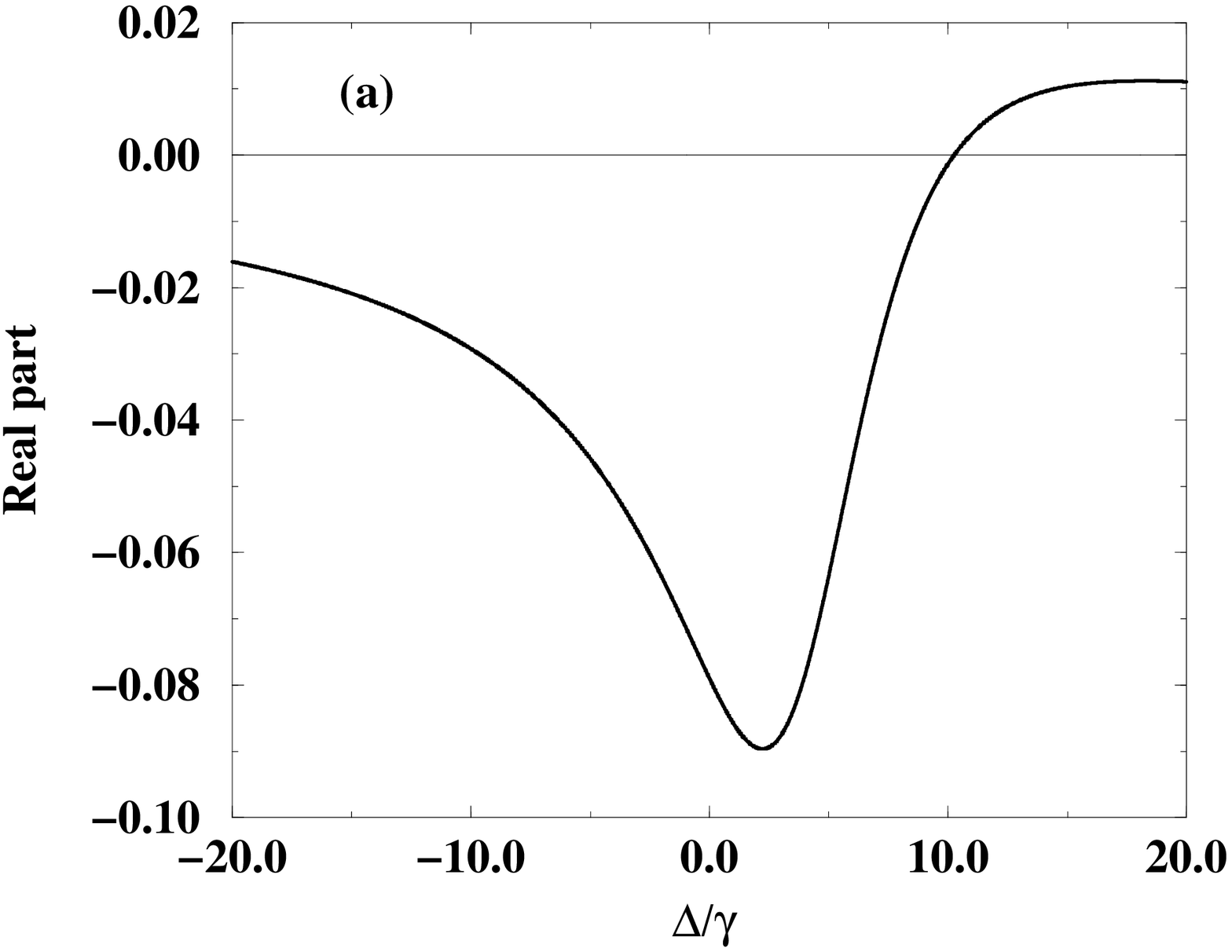}}&
\scalebox{0.37}{\includegraphics{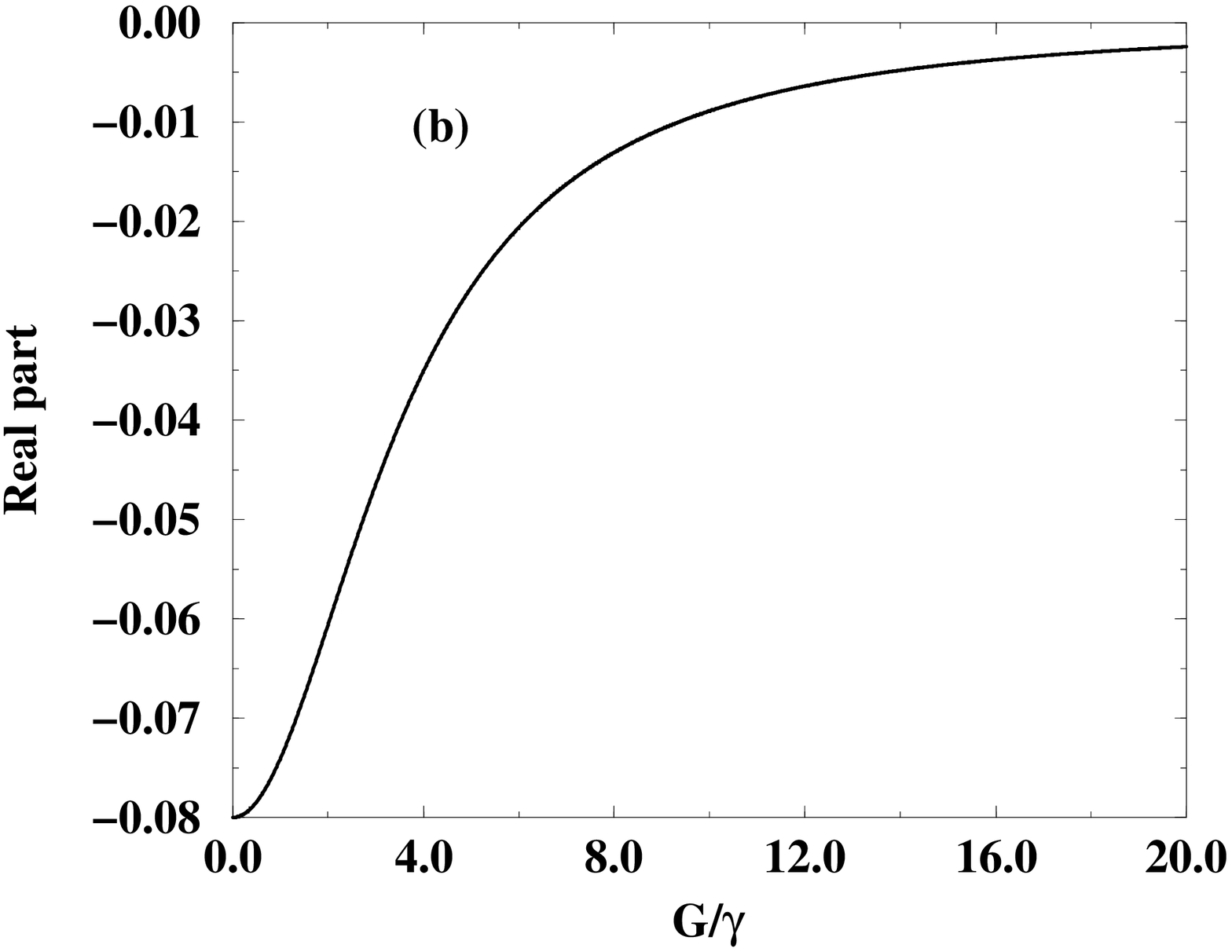}}
\end{tabular}
\caption{\label{fig3}Variation of real part of $\chi_-$ in unit of
$N|\vec{d}_{e_+g_-}|^2/\hbar\gamma$ at two-photon resonance
($\delta=\Delta$) (a) with respect to $\Delta$ for fixed
$G=0.5\gamma$ and (b) with respect to $G$ for fixed $\Delta=B'-B$.
The other parameters are the same as in Fig.~\ref{chap5fig2}.}
\end{figure*}
The relevant zeroth order coherence terms turn out to be
\begin{subequations}
\begin{eqnarray}
\label{rhogeEx}\tilde{\rho}_{g_+e_+}^{(0)}&=&-\frac{ixG^*}{cQ}(\gamma_1+\gamma_2)\;,\\
\tilde{\rho}_{g_-e_-}^{(0)}&=&-\frac{iyG^*}{dQ}(\gamma_1+\gamma_2)\;,
\end{eqnarray}
\end{subequations}
which vanish in absence of any control field (i.e., for $G=0$).

The susceptibilities of the $\sigma_\mp$ components of the probe
field can be obtained from first-order solutions which we write as
\begin{eqnarray}
\tilde{\rho}_{e_+g_-}^{'(-1)}&=&\frac{1}{M_1}\left[Ga_+p_+\tilde{\rho}_{g_+e_+}^{(0)}+Gb_+p_+\tilde{\rho}_{g_-e_-}^{(0)}\right.\nonumber\\
&+&\left.i\{a_+b_+p_++|G|^2(a_++b_+)\}(\tilde{\rho}_{g_-g_-}^{(0)}-\tilde{\rho}_{e_+e_+}^{(0)})\right]\;,\nonumber\\
&&\label{rho_egN}\\
\tilde{\rho}_{e_-g_+}^{'(+1)}&=&\frac{1}{M_2}\left[Gb_-q_-\tilde{\rho}_{g_+e_+}^{(0)}+Ga_-q_-\tilde{\rho}_{g_-e_-}^{(0)}\right.\nonumber\\
&+&\left.i\{a_-b_-q_-+|G|^2(a_-+b_-)\}(\tilde{\rho}_{g_+g_+}^{(0)}-\tilde{\rho}_{e_-e_-}^{(0)})\right]\;,\nonumber\\
&&
\end{eqnarray}
where
\begin{eqnarray}
M_1&=&a_+b_+p_+q_++|G|^2(p_++q_+)(a_++b_+)\;,\nonumber\\
M_2&=&a_-b_-p_-q_-+|G|^2(p_-+q_-)(a_-+b_-)\;,\nonumber
\end{eqnarray}
and
\begin{eqnarray}
a_\pm &=& -i\omega_{12}\pm 2iB +\Gamma_{e_+e_-}\;,\nonumber\\
b_\pm &=& -i\omega_{12}\pm 2iB'+\Gamma_{g_+g_-}\;,\nonumber\\
p_\pm &=& -i\omega_{12} \pm i(\Delta+2B)+\Gamma\;,\nonumber\\
q_\pm &=& -i\omega_{12} \pm i(-\Delta+2B')+\Gamma\;.\nonumber
\end{eqnarray}
Note that the difference between the frequencies of the probe and
control fields is given by $\omega_{12}=\delta-\Delta+2B'$. The
above susceptibility is to be compared with the one in the absence
of cross-talk (Case I). For the Case I, we have
\begin{subequations}
\begin{eqnarray}
\label{rho_eg1}\tilde{\rho}_{e_+g_-}^{'(-1)}&=&\frac{ib_+}{b_+q_++|G|^2}(\tilde{\rho}_{g_-g_-}^{(0)}-\tilde{\rho}_{e_+e_+}^{(0)})\\
\label{cohG1}&=&\frac{-i\{i(\delta-\Delta)-\Gamma_{g_+g_-}\}}{\{i(\delta-\Delta)-\Gamma_{g_+g_-}\}(i\delta-\Gamma)+|G|^2}\;,
\end{eqnarray}
\end{subequations}
where we have used the fact that in steady state,
$\tilde{\rho}_{g_-g_-}^{(0)}=1$ and the populations in the other
levels vanish. In this case, the zeroth-order coherence between
the levels $|e_+\rangle, |g_+\rangle$ also vanishes. From
Eq.~(\ref{cohG1}), we find that the real part of the
susceptibility vanishes at the detunings $\delta=\Delta+\Delta_-$,
satisfying the following equation:
\begin{equation}
\Delta_-^3+\Delta\Delta_-^2+(\Gamma_{g_+g_-}^2+2\Gamma\Gamma_{g_+g_-}+|G|^2)\Delta_-+\Delta\Gamma_{g_+g_-}^2=0\;.
\end{equation}
Three different solutions of the above equation for $\Delta_-$
corresponding to vanishing real part of $\chi_-$, can be obtained
from the Cardano's formula \cite{cardano}, given by
\begin{equation}
\Delta_-^1=-\frac{\Delta}{3}+A_+\;,
\Delta_-^{2,3}=-\frac{\Delta}{3}-\frac{1}{2}A_+\pm\frac{i}{2}\sqrt{3}A_-\;,
\end{equation}
where
\begin{eqnarray}
&&A_\pm=[R+\sqrt{Q^3+R^2}]^{1/3}\pm
[R-\sqrt{Q^3+R^2}]^{1/3}\;,\nonumber\\
&&Q=\frac{1}{9}(3a_1-\Delta^2)\;, R=\frac{1}{54}(9\Delta
a_1-27a_0-2\Delta^3)\;,\nonumber\\
&&a_0=\Delta\Gamma_{g_+g_-}^2\;,
a_1=\Gamma_{g_+g_-}^2+2\Gamma\Gamma_{g_+g_-}+|G|^2\;.
\end{eqnarray}

\section{\label{sec:absorb}Absorption and dispersion profiles}

We first recall the features of the $\Lambda$ system of
Fig.~\ref{fignew}(b). The usual line-shapes can be obtained for a
resonant control field, i.e., by putting $\Delta=0$ in the
susceptibility given by Eq.~(\ref{cohG1}). Then the absorption and
dispersion profiles would be symmetric around $\delta=0$. However,
for non-zero $\Delta$, the line-shapes depend upon the values of
$\Delta$. We show the dispersion and absorption profiles of the
$\sigma_-$ component in Figs.~\ref{chap5fig2}(a) and
\ref{chap5fig2}(b) for a fixed detuning $\Delta=B'-B$ of the
control field. Clearly, the real and imaginary parts of the
susceptibility $\chi_-$ [$\equiv
(N|\vec{d}_{e_+g_-}|^2/\hbar\gamma)\tilde{\rho}_{e_+g_-}^{'(-1)}$,
$N$ being the number density of the atomic medium] vanish at the
two-photon resonance $\delta=\Delta$, which occurs at
$\delta=\Delta=4\gamma$ for $B=2\gamma$. These are the usual
features of a $\Lambda$ system at two-photon resonance.

Next we analyze the four-level system  of Fig.~\ref{chap5fig1}. We
show the dispersion and absorption profiles of the $\sigma_-$
component in Figs.~\ref{chap5fig2} for $\Delta=B'-B$
\cite{sigmaplus}.
\begin{figure}
\scalebox{0.37}{\includegraphics{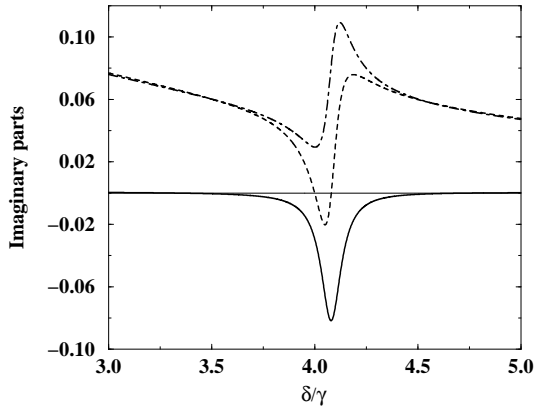}}
\caption{\label{fig4}Variation of imaginary parts of the first
term (solid line) and the third term (dot-dashed line) of
Eq.~(\ref{rho_egN}) and of $\tilde{\rho}_{e_+g_-}^{'(-1)}$ (dashed
line) with detuning $\delta$ of the probe field. We have chosen
$G=0.5\gamma$ and the other parameters are the same as in
Fig.~\ref{chap5fig2}.}
\end{figure}
\begin{figure}
\scalebox{0.37}{\includegraphics{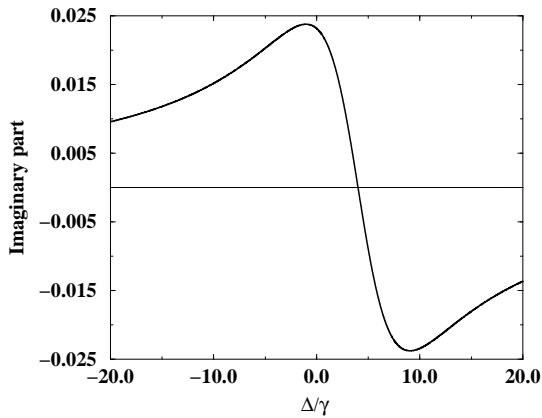}}
\caption{\label{fig5}Variation of imaginary part of $\chi_-$ in
unit of $N|\vec{d}_{e_+g_-}|^2/\hbar\gamma$ with $\Delta=\delta$.
The parameters chosen here are the same as in Fig.~\ref{fig3}(a).}
\end{figure}
At two-photon resonance (i.e., at $\delta=\Delta$), the real part
of the susceptibility $\chi_-$ is {\it non-zero and negative\/} in
contrast to the case of a $\Lambda$ system. On the other hand, at
two-photon resonance $\delta=\Delta$, the medium continues to
remain transparent as in case of a $\Lambda$-system. Further, at
certain region of the detuning $\delta$ ($> \Delta$) of the probe
field, the imaginary part of $\chi_-$ becomes {\it negative\/},
leading to the {\it gain in the $\sigma_-$ component\/}. For the
case of a $\Lambda$-system, there would be no possibility of gain
in the medium [solid line in Fig.~\ref{chap5fig2}(b)] for any
$\delta$. In the next section, we analyze these new features in
terms of the new coherence that we discussed in the Sec.~II.

We should mention here that two-photon gain in hot alkali vapor
has been reported in recent experiments \cite{gau1}. The gain
involves absorption of two photons from the detuned $\sigma_-$
polarized control field and emission of two photons in
$\pi$-polarized probe field at Raman resonance. Detailed
theoretical analysis \cite{gau2} has shown that these are due to
quantum interference of several excitation pathways, originating
from hyperfine structures. Because in our model, gain arises due
to interference of two different pathways, as described Sec. II,
we should emphasize the main difference between the model in
\cite{gau2} and our model. We consider a different set of
polarizations and frequency of the electric fields interacting
with the non-degenerate electronic levels, contrary to
\cite{gau2}, where degenerate hyperfine levels have been
considered. In our model, the gain, associated with emission of a
single photon in $\sigma_-$ component, is essentially a two-photon
process, and thus much larger in effect compared to that arising
due to four-photon process described in \cite{gau1,gau2}. In
addition, gain occurs when the probe field is {\it not\/} at Raman
resonance with the control field.

\section{\label{sec:discuss}Discussions}
\subsection{Origin of non-zero susceptibility}
We start with the expression (\ref{rho_eg1}) for
$\tilde{\rho}_{e_+g_-}^{'(-1)}$ in the $\Lambda$-system. Clearly,
the susceptibility of the probe field arises only by the
population difference between the relevant levels $|e_+\rangle$
and $|g_-\rangle$, as the zeroth-order coherence between the
levels $|g_+\rangle$ and $|e_+\rangle$ is zero in this case. In
addition, at two-photon resonance ($\delta=\Delta$), $b_+=0$ and
the susceptibility vanishes. But, in case of the four-level system
of Fig.~\ref{chap5fig1}, the coherence $\tilde{\rho}_{g_+
e_+}^{(0)}$ also contributes to the susceptibility $\chi_-$ [see
Eq. (\ref{rho_egN})], while contribution from
$\tilde{\rho}_{g_-e_-}^{(0)}$ to $\chi_-$ vanishes at two-photon
resonance ($\delta=\Delta$) as $b_+=0$. Then, using
Eqs.~(\ref{popul}) and (\ref{rhogeEx}), we can write for
$\delta=\Delta=B'-B$
\begin{equation}
\tilde{\rho}_{e_+g_-}^{'(-1)}=\frac{ix(\gamma_1+\gamma_2)}{2q_+Q}\left(1-\frac{q_+}{c}\right)\;,
\end{equation}
where $p_+=q_+=c^*=\Gamma+i(B-B')$, and $Q>0$. The first term
inside the bracket in the above equation is due to the
contribution of
$\tilde{\rho}_{g_-g_-}^{(0)}-\tilde{\rho}_{e_+e_+}^{(0)}$ and the
second term is due to the coherence $\tilde{\rho}_{g_+e_+}^{(0)}$.
We see from the above expression, that real parts of these two
terms cancel each other, and it is essentially their imaginary
parts which contribute to the susceptibility. We find that
\begin{equation}
\label{nonzero}\tilde{\rho}_{e_+g_-}^{'(-1)}=\frac{-(B'-B)}{2\left\{\Gamma^2+(B'-B)^2+2|G|^2\right\}}
\end{equation}
which is non-zero as $B'\neq B$. Thus nonzero susceptibility of
the system of Fig.~\ref{chap5fig1} manifests itself as an effect
of the {\it zeroth order coherence\/} in the
$|e_+\rangle\leftrightarrow |g_+\rangle$ transition. Further, it
is essentially associated with no absorption, as
Eq.~(\ref{nonzero}) has no imaginary part. We should mention here
that the susceptibility becomes negative due to larger Land\'e $g$
factor of the ground state manifold (i.e., $B'>B$). Clearly, in
absence of the magnetic field, the medium becomes isotropic, and
no special features can be found in the dispersion profile.

In Fig.~\ref{fig3}(a), we show how the real part of $\chi_-$
varies with the detuning $\Delta$ of the control field at
two-photon resonance $\delta=\Delta$. Clearly, even for resonant
control field (i.e., for $\Delta=0$), the susceptibility is
negative. Moreover, the real part of $\chi_-$ becomes zero at
two-photon resonance for certain value of $\delta$. Putting
$b_+=0$ in Eq.~(\ref{rho_egN}), one can calculate the
corresponding value
\begin{equation}
\delta_0=\frac{2(B'-B)^2+\Gamma^2}{B'-B}\;.
\end{equation}
For the present parameters, we find that $\delta_0=10.25\gamma$,
which is shown in Fig.~\ref{fig3}(a). In fact, for
$\delta=\Delta=\delta_0$, the contributions of the population
difference term and the coherence term to Re($\chi_-$) cancel each
other and the susceptibility of the probe field vanishes. In
Fig.~\ref{fig3}(b), we show that the control field can be a good
control parameter for the susceptibility. The susceptibility
remains negative for the entire range of $G$ at two-photon
resonance. This is unlike the case of a $\Lambda$ system [see
Fig.~\ref{fignew}(b)] for which the real part of $\chi_-$ would
remain zero at two-photon resonance irrespective of the value of
Rabi frequency $G$ of the control field .

\subsection{Origin of gain}
It is well understood that, the absorption spectrum of the probe
field in the $\Lambda$ system of Fig.~\ref{fignew}(b) shows
Autler-Townes doublet. Moreover, there does not arise any gain in
the medium. At two-photon resonance $\delta=\Delta$, the
absorption becomes zero. However, the case of the four-level
system of Fig.~\ref{chap5fig1} is different. We already have noted
that the susceptibility $\chi_-$ of the $\sigma_-$ component is
contributed by two terms : $\tilde{\rho}_{g_+e_+}^{(0)}$ and
$\tilde{\rho}_{g_-g_-}^{(0)}-\tilde{\rho}_{e_+e_+}^{(0)}$. We show
the individual contribution of these two terms in the absorption
spectra in Fig.~\ref{fig4}. From this figure one can see that at
certain region of frequency ($\delta > B'-B$), the negative
contribution of $\tilde{\rho}_{g_+e_+}^{(0)}$ is larger in
magnitude than the positive contribution of
$\tilde{\rho}_{g_-g_-}^{(0)}-\tilde{\rho}_{e_+e_+}^{(0)}$. Thus,
gain arises in the medium. In our model, this novel feature can be
attributed to the control field $G$ which gives rise to the
non-zero coherence $\tilde{\rho}_{g_+e_+}^{(0)}$. Further, from
the expressions of $\tilde{\rho}^{(0)}_{g_\pm g_\pm}$ and
$\tilde{\rho}^{(0)}_{e_\pm e_\pm}$ [Eqs.~(\ref{popul})], one sees
that the zeroth-order populations in both the $|g_\pm\rangle$
levels are larger than those in the levels $|e_\pm\rangle$ due to
the presence of non-zero decay terms $\gamma_1$ and $\gamma_2$.
Thus there is no population inversion in bare basis and we have
{\it gain without inversion\/}. Note that at two-photon resonance
($\delta=\Delta$), the contributions from the terms
$\tilde{\rho}_{g_+e_+}^{(0)}$ and
$\tilde{\rho}_{g_-g_-}^{(0)}-\tilde{\rho}_{e_+e_+}^{(0)}$ to the
absorption profile cancel each other, leading to transparency. We
should further mention here that the contribution of
$\tilde{\rho}_{g_-e_-}^{(0)}$ to the gain is negligible for all
$\delta$.


\section{conclusions}
In conclusion, we have shown how a control field can give rise to
new coherence effects in a specific four-level system. The control
field leads to multiple pathways, interference between which leads
to the effects like gain without inversion and non-zero
susceptibility associated with zero absorption. We have explained
these results in terms of the new coherence arising due to this
interference.

\begin{acknowledgements}
I am very much thankful to Prof. G. S. Agarwal for his valuable
suggestions and discussions during this work.
\end{acknowledgements}


\begin{thebibliography}{99}
\bibitem{note}
It is assumed here that there is no spontaneous decay from the
ground levels $|g_\pm\rangle$.
\bibitem{sigmaplus}
The similar profiles of $\sigma_+$ component could also be
considered. However in terms of physical origin, the profiles of
$\sigma_\pm$ components are complementary to each other. So it
suffices to consider the either component.
\end{thebibliography}
\end{document}